# Simulation of non-resonant stellarator divertor

Alkesh Punjabi 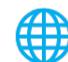, and Allen H. Boozer

**COLLECTIONS**

This paper was selected as an Editor's Pick

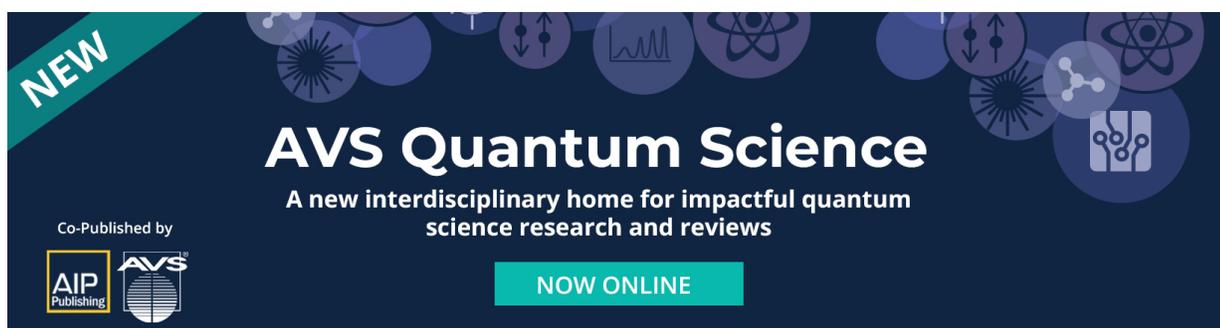 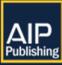
View Online        Export Citation        CrossMark

**ARTICLES YOU MAY BE INTERESTED IN**

A tokamak pertinent analytic equilibrium with plasma flow of arbitrary direction
Physics of Plasmas **26**, 124501 (2019); https://doi.org/10.1063/1.5120341

Excitation of VLF perturbations in the F-region of the ionosphere by beating HF O-mode waves
Physics of Plasmas **27**, 012101 (2020); https://doi.org/10.1063/1.5119514

# Simulation of non-resonant stellarator divertor

Alkesh Punjabi[1,a)] and Allen H. Boozer[2,b)]

**AFFILIATIONS**

[1]Department of Mathematics, Hampton University, Hampton, Virginia 23668, USA
[2]Department of Applied Physics and Applied Mathematics, Columbia University, New York, New York 10027, USA

a)alkesh.punjabi@hamptonu.edu
b)ahb17@columbia.edu

**ABSTRACT**

An efficient numerical method of studying nonresonant stellarator divertors was introduced in Boozer and Punjabi [Phys. Plasmas 25, 092505 (2018)]. This method is used in this paper to study a different magnetic field model of a nonresonant divertor. The most novel and interesting finding of this study is that diffusive magnetic field lines can be distinguished from lines that exit through the primary and the secondary turnstile, and that below some diffusive velocity, all lines exit through only the primary turnstile. The footprints of each family are stellarator symmetric and have a fixed location on the wall for all velocities. The probability exponent of the primary turnstile is $d_1 = 9/4$ and that of the secondary turnstile is $d_2 = -3/2$. This study also addresses the issues of an inadequate separation of the chamber walls from the outermost confining magnetic surface and a marginal step size of the numerical integrations that could compromise the interpretation of the earlier results [Boozer and Punjabi, Phys. Plasmas 25, 092505 (2018)]. The previous value of $d_1 = 2$ is within the error bar of $d_1 = 9/4$ estimated here.

Published under license by AIP Publishing. https://doi.org/10.1063/1.5113907

## I. INTRODUCTION

The divertors in stellarators are of two types: resonant and nonresonant.

In resonant stellarator divertors, an external magnetic perturbation, which is resonant on a magnetic surface in the edge, is applied to create magnetic islands outside the last confining magnetic surface. The resonant perturbation destroys the magnetic surface and creates magnetic islands in its place. The islands divert the plasma exhaust to walls where it is pumped away. The W7-X stellarator has a resonant divertor on the surface with rotational transform $\iota = 5/5$ with five islands in the poloidal plane.[1] The advantages of the resonant divertor are that the divertor region is compact, located very close to the body of the plasma, and the divertor has a long connection length. Resonant divertors also have some disadvantages. A fixed resonant rotational transform has to be maintained at the plasma edge to form islands. This is difficult when the plasma evolution naturally has a large variation in the net toroidal current. The structures that collect the diverted field lines are separated by only a small distance from the main plasma. This is despite the many toroidal turns that the field lines make in going from the outermost confining surface to the collecting surfaces.

A nonresonant divertor is produced by using the external magnetic field to force a sharp edge on the plasma surface. Sharp edges are a requirement for a nonresonant divertor. The sharp edges on the outermost confining surface are like X-points in tokamaks. However, the separatrices of these X-points in nonresonant divertors do not cover the whole toroidal angle; they only partially cover the toroidal angle. Nonresonant divertors can be insensitive to the net plasma current, and the diverted field lines can intersect the divertor structures far from the main plasma body. The general geometry of the magnetic field lines controls where the flux tubes filled with plasma strike the walls. The flux tubes tend to have a robust strike location on walls independent of plasma details.[2] Nonresonant divertors have been discussed for quasiaxisymmetric stellarators in Refs. 3 and 4. Helically Symmetric Experiment (HSX) is an example of a resilient nonresonant divertor.[2,5]

Nonresonant divertors are understudied in stellarator physics and characterization, both theoretically and the numerically, is extremely useful to advance the stellarator concept. As such, the topic is important for the field. This paper is an exploration of the behavior of a nonresonant divertor for a one period stellarator. This paper is a follow-up of our previous work.[6]

In nonaxisymmetric toroidal fusion plasmas, the region outside the outermost confining magnetic surface has one or more families of magnetic cantori, islands, and stochastic field lines. Each family of cantori has its magnetic turnstile. A magnetic turnstile is a double flux tube; in one part, field lines move outwards, and in another adjacent part, field lines move inwards. The incoming and the outgoing magnetic fluxes are exactly equal. Magnetic field lines just outside the outermost good surface reach the wall through the turnstiles.[6–8] The cantori family closest to the outermost confining surface is called the first or the primary family and its turnstile is called primary turnstile. Further out there can be a second family of cantori and its turnstile is called secondary turnstile. The points where the field lines coming through the primary turnstile strike the wall are called the primary footprint, and the points where the field lines coming through the secondary turnstile strike the wall are called the secondary footprint.

Recently, an efficient simulation method was developed for carrying out topological studies of stellarator divertors. This method is described in Ref. 6, and illustrative examples were given. Here, we use this method to carry out additional studies of nonresonant stellarator divertors.

In our previous paper[6] on the simulation of stellarator divertors, some illustrative results from the simulation of nonresonant stellarator divertors were presented. The simulation in Ref. 6 was for all five periods of the stellarators with a step-size of $\delta\varphi = 2\pi/360$ for the symplectic integration of magnetic field line trajectories over 10 000 toroidal circuits with the shape parameters $\varepsilon_0 = 1/2$, $\varepsilon_t = 1/2$, and $\varepsilon_x = -1/5$. The footpoints in periods 2–5 were shifted to the first period of the stellarator and coalesced into a single footprint. This paper is on the simulation of the nonresonant stellarator of a single-period of the stellarator with a step-size of $\delta\zeta = 2\pi/3600$ over 100 000 toroidal circuits of the single period with the shape parameters $\varepsilon_0 = 1/2$, $\varepsilon_t = 1/2$, and $\varepsilon_x = -0.31$. $\varphi$ is the toroidal angle of the stellarator and $\zeta$ is the toroidal angle of a single period. $\varphi$ and $\zeta$ are related by $\zeta = n_p \varphi$, where $n_p$ is the number of periods of the stellarator. In this paper, all five periods of the stellarator are considered to be strictly identical. The step-size for the single-period simulation here is effectively fifty times smaller than the five-period simulation. The longer integration from 10 000 toroidal circuits of the stellarator to 100 000 toroidal circuits of a single period effectively doubles the integration time because 100 000 transits of a single-period are double the 10 000 transits of a 5-period stellarator. The wall is moved further out to allow for a fuller exploration of the magnetic topology between the outermost confining surface and the wall.

This paper is organized as follows: Sec. II describes the simulation methodology used to study the nonresonant stellarator divertor. Magnetic fields are divergence free and their field lines cannot spiral outwards. Nevertheless, the structures formed by magnetic field lines can be studied without prejudice by introducing a nonzero outward spiraling of the lines. This spiraling is described as a radial velocity, the outward motion in the toroidal flux per toroidal transit. Section III gives the results of the simulation. Section IV gives the summary of results and the conclusions.

The results reported in this paper differ in three ways from our earlier publication:[6] (1) the wall, which is used to represent the divertor strike points, has been moved much further back from the outermost confining magnetic surface. The outermost confining magnetic surface in our earlier results was found to contact the wall over a short range of toroidal angles. (2) The step size of the symplectic integration has been made fifty times smaller than before. The magnetic fields that are being studied have the stellarator symmetry of all standard stellarator designs.[9] When the integration step size is too large, the results can break this symmetry. Finite step-size effects that break stellarator symmetry are now unobservable. (3) A different model of the magnetic field of a nonresonant divertor is used. Equation (6) gives a power-law approximation to the probability that particles on a cantorus outside the outermost confining magnetic surface will enter a magnetic flux tube that goes to the wall. The results remain consistent with this power-law form. The power-law statement is an approximation, and the phenomena associated with turnstiles are too complicated for it to be an exact result. The previous value of two for the probability exponent for the primary cantori is found to be within the error bar of the value that is estimated here.

## II. SIMULATION METHOD FOR THE NONRESONANT STELLARTOR DIVERTOR

In this section, we describe the method we have used for the simulation in a single period of the nonresonant stellarator divertor.

### A. Hamiltonian for the field line trajectories in the nonresonant stellarator divertor

The magnetic field $\vec{B}$ in toroidal fusion plasmas has generalized contravariant representation as

$$\vec{B} = \vec{\nabla}\psi_t \times \vec{\nabla}\theta + \vec{\nabla}\varphi \times \vec{\nabla}\psi_p(\psi_t, \theta, \varphi).$$

This is called canonical representation of magnetic field.[10] $\theta$ is the poloidal angle, $\varphi$ is the toroidal angle, $\psi_t$ is the toroidal magnetic flux, and $\psi_p$ is the poloidal magnetic flux. The poloidal flux is written using $(\psi_t, \theta, \varphi)$ as coordinates. This is a valid choice if $\vec{B} \cdot \vec{\nabla}\varphi = (\vec{\nabla}\psi_t \times \vec{\nabla}\theta) \cdot \vec{\nabla}\varphi$ is nonzero. This is the case in a stellarator. The magnetic field lines are given by the equations

$$d\psi_t/d\varphi = \vec{B} \cdot \vec{\nabla}\psi_t / \vec{B} \cdot \vec{\nabla}\varphi$$

and

$$d\theta/d\varphi = \vec{B} \cdot \vec{\nabla}\theta / \vec{B} \cdot \vec{\nabla}\varphi,$$

which can be rewritten using the canonical representation as

$$d\psi_t/d\varphi = -d\psi_p/d\theta$$

and

$$d\theta/d\varphi = d\psi_p/d\psi_t.$$

These equations are mathematically identical to the Hamiltonian equations with Hamiltonian $H(q, p, t)$ that is periodic in the time like variable. Hamiltonian of this form is said to have one and one-half degrees of freedom and is well-known as the simplest Hamiltonian system in which chaotic trajectories can arise. We study a five period, $n_p = 5$, stellarator and assume that all five periods are exactly identical. We calculate the field line trajectories in a single period. The Hamiltonian for the trajectories of magnetic field lines in a single period is given by[6]

$$\frac{\psi_p}{\bar{\psi}_g} = \left[\iota_0 + \frac{\varepsilon_0}{4}((2\iota_0 - 1)\cos(2\theta - \zeta) + 2\iota_0 \cos 2\theta)\right]\left(\frac{\psi_t}{\bar{\psi}_g}\right)$$
$$+ \frac{\varepsilon_x}{8}[(4\iota_0 - 1)\cos(4\theta - \zeta) + 4\iota_0 \cos 4\theta]\left(\frac{\psi_t}{\bar{\psi}_g}\right)^2$$
$$+ \frac{\varepsilon_t}{6}[(3\iota_0 - 1)\cos(3\theta - \zeta) - 3\iota_0 \cos 3\theta]\left(\frac{\psi_t}{\bar{\psi}_g}\right)^{3/2}, \quad (1)$$

where $\psi_p$ is the poloidal flux, $\psi_t$ is the toroidal flux, $\iota_0$ is the rotational transform on the magnetic axis, and $\zeta$ is the toroidal angle of the single period. $\zeta$ is related to the toroidal angle $\varphi$ by $\zeta = n_p \varphi$. $\theta$ is the poloidal angle. Radial position $r$ is given by $r \equiv \sqrt{\psi_t/\pi B_c}$. $B_c$ is a characteristic magnetic field strength. The poloidal flux $\psi_p$ is per period. The rotational transform per period was chosen to be $\iota_0 \cong 0.15$. The shape parameters $\varepsilon_0, \varepsilon_t$, and $\varepsilon_x$ control the elongation, triangularity, and the sharpness of the edges of the outermost confining surface in the nonresonant stellarator divertor, respectively. We have chosen $\varepsilon_0 = \varepsilon_t = 1/2$, and $\varepsilon_x = -0.31$. $\bar{\psi}_g$ is an averaged toroidal flux used to normalize magnetic fluxes.

### B. Map equations for the field line trajectories in the nonresonant stellarator divertor

Let $\bar{\psi}_t \equiv \psi_t/\bar{\psi}_g$ and $\bar{\psi}_p \equiv \psi_p/\bar{\psi}_g$. Then, the area-preserving map equations for the trajectories of magnetic field line trajectories in a single period of nonresonant stellarator are

$$\psi_t^{(j+1)} = \psi_t^{(j)} - \frac{\partial \psi_p\left(\psi_t^{(j+1)}, \theta^{(j)}, \zeta^{(j)}\right)}{\partial \theta^{(j)}} \delta\zeta, \quad (2)$$

$$\theta^{(j+1)} = \theta^{(j)} + \frac{\partial \psi_p\left(\psi_t^{(j+1)}, \theta^{(j)}, \zeta^{(j)}\right)}{\partial \psi_t^{(j+1)}} \delta\zeta, \quad (3)$$

$$\zeta^{(j+1)} = \zeta^{(j)} + \delta\zeta. \quad (4)$$

$\delta\zeta = 2\pi/3600$ radian is the step-size of the map. For the forward moving trajectories,

$$\left(\psi_t^{(j)}, \theta^{(j)}, \zeta^{(j)}\right) \to \left(\psi_t^{(j+1)}, \theta^{(j+1)}, \zeta^{(j+1)}\right).$$

Equation (2) is solved for $\psi_t^{(j+1)}$; then, Eqs. (3) and (4) are solved for $\theta^{(j+1)}$ and $\zeta^{(j+1)}$. Equation (2) for $\psi_t^{(j+1)}$ is the implicit equation and equations for $\theta^{(j+1)}$ and $\zeta^{(j+1)}$ are explicit. For the backward trajectories,

$$\left(\psi_t^{(j+1)}, \theta^{(j+1)}, \zeta^{(j+1)}\right) \to \left(\psi_t^{(j)}, \theta^{(j)}, \zeta^{(j)}\right),$$

the map equations are solved in the reverse order. For the backward trajectories, Eq. (3) is implicit and Eqs. (2) and (4) are explicit. To calculate the strike points on the wall, the continuous analogs of the map equations are used. Continuous analogs of the forward and backward map equations are also area-preserving.

### C. Spiraling radial velocity[6]

Field lines are given a constant radial velocity in $\psi_t$-space. The radial velocity is denoted by $u_\psi$. The radial velocity causes the lines to spiral outwards. The radial velocity allows the exploration of magnetic topology in the region between the outermost confining surface and the wall. The Hamiltonian equation $d\psi_t/d\zeta = -\partial \psi_p/\partial \theta$ is modified to $d\psi_t/d\zeta = -\partial \psi_p/\partial \theta + u_\psi$. This is implemented in the map equations by $\psi_t \to \psi_t + u_\psi \delta\zeta$ after each iteration of the map.

### D. Model for loss of field lines to walls through magnetic turnstiles[6]

The model used to describe the loss of magnetic field lines to the wall through magnetic turnstiles is based on the probability $P_t(\bar{\psi}_t)$.

$P_t(\bar{\psi}_t)$ is defined as the probability per radian advance in the toroidal angle $\varphi$ that a magnetic field line will be lost by passing through a turnstile. The probability is assumed to have a power-law form when $\bar{\psi}_t > \bar{\psi}_o$,[6]

$$P_t(\bar{\psi}_t) = (d+1)c_p \left(\frac{\bar{\psi}_t - \bar{\psi}_o}{\bar{\psi}_o}\right)^d, \quad (5)$$

where $\bar{\psi}_o$ is the average toroidal flux of the good surface on which lines start.

The probability is characterized by two constants: the first is its strength $c_p$ and the second is the power $d$ with which it opens with distance.

It is possible that additional turnstiles can arise on quasisurfaces at $\bar{\psi}_t = \bar{\psi}_i$, and then, $P_t(\bar{\psi}_t)$ can be written as

$$P_t(\bar{\psi}_t) = \sum_i (d_i + 1)c_i \left(\frac{\bar{\psi}_t - \bar{\psi}_i}{\bar{\psi}_i}\right)^{d_i} H(\bar{\psi}_t - \bar{\psi}_i), \quad (6)$$

where the Heaviside step function is $H(x \geq 0) = 1$ and $H(x < 0) = 0$. The scaling is said to be universal if all the powers $d_i$ are the same.[6,11] The width $w$, loss-time $\zeta_\ell$, and flux decay $N_B(\zeta)/N_0$ scale, respectively,[6] as

$$w \equiv \left(u_\psi/c_p\right)^{\frac{1}{d+1}}, \quad (7)$$

$$\zeta_\ell \equiv w/u_\psi = (1/c_p)(c_p/u_\psi)^{\frac{d}{d+1}}, \quad (8)$$

and

$$N_B(\zeta) = N_0 \exp\left[-\left(\frac{\zeta - \zeta_o}{\zeta_\ell}\right)^{d+1}\right] \text{ for } \zeta > \zeta_o. \quad (9)$$

$N_0$ is the number of lines that start on a good surface, $N_B(\zeta)$ is the number of lines remaining at toroidal angle $\zeta$, and $\zeta_0$ is the toroidal angle when the first line is lost to the wall.

### E. Calculation of the width of footprints

The width of the footpoints $\delta\theta$ about a smooth curve on the wall gives a good estimate of the width of the split in the last good surface given by a turnstile. The coordinate system used to describe the wall is

$$\vec{x}_w(\theta, \zeta) = (R_w + b\cos\theta)\hat{R}(\zeta) - b\sin\theta\hat{Z},$$

where $\hat{R}(\zeta)$, $\hat{\zeta} = d\hat{R}/d\zeta$, and $\hat{Z}$ are the three unit vectors of cylindrical coordinates. $R_w$ is the major radius of the wall and $b$ is the minor radius. The curve on the wall around which the footpoints lie is denoted by $\theta_c(\zeta)$. The aspect ratio of the wall will be assumed to be very large $R_w/b \to \infty$. The separation of the $i$th footpoint from a given curve is

$$\vec{\Delta}_i \equiv \vec{x}_w(\theta_i, \zeta_i) - \vec{x}_w(\theta_c(\zeta), \zeta),$$

where $\zeta$ is to be chosen for each particle to minimize $\Delta_i$. In the limit $R_w/b \to \infty$, this choice is $\zeta = \zeta_i$. Assuming that the width of the footpoints about the curve $\delta\theta$ is small compared to unity, let $\theta_i = \theta_c(\zeta_i) + \delta\theta_i$, then

$$\vec{\Delta}_i \equiv \vec{x}_w(\theta_c(\zeta_i) + \delta\theta_i, \zeta_i) - \vec{x}_w(\theta_c(\zeta_i), \zeta_i)$$
$$= \left(\frac{\partial x_w}{\partial \theta}\right)_{\theta_c} \delta\theta_i.$$

The magnitude of the separation is then easily shown to be $\Delta_i = b\delta\theta_i$.

The curve in the wall along which footpoints lie will be written as

$$\theta_c(\zeta) = \theta_0 + \iota_w \zeta + \kappa \sin \zeta, \quad (10)$$

where the three constants have the interpretation that $\theta_0$ is the location of the curve a $\zeta = 0$, $\iota_w$ is an effective transform in the wall, and $\kappa$ is nonzero if the curve is not straight. In principle, this is the first of an arbitrarily large number of Fourier terms. The constants describing the curve should be chosen to minimize the scatter. An approximate way to do this is to force the first three moments of $\delta\theta_i \equiv \theta_i - \theta_c(\zeta_i)$ vanish when averaged over the $I$ footpoints. The three equations that give the three constants are

$$\langle \theta_i \rangle \equiv \frac{1}{I} \sum_{i=1}^{I} \theta_i,$$
$$\langle \theta_i \rangle = \theta_0 + \iota_w \langle \zeta_i \rangle + \kappa \langle \sin \zeta_i \rangle,$$
$$\langle \theta_i \sin \zeta_i \rangle = \theta_0 \langle \sin \zeta_i \rangle + \iota_w \langle \zeta_i \sin \zeta_i \rangle + \kappa \langle \sin^2 \zeta_i \rangle,$$
$$\langle \theta_i \sin^2 \zeta_i \rangle = \theta_0 \langle \sin^2 \zeta_i \rangle + \iota_w \langle \zeta_i \sin^2 \zeta_i \rangle + \kappa \langle \sin^3 \zeta_i \rangle.$$

The width of the points near the curve $\delta\theta$ is defined by

$$\delta\theta \equiv 2\sqrt{\langle \delta\theta_i^2 \rangle}, \quad (11)$$

where $\langle \delta\theta_i \rangle^2 / \langle \delta\theta_i^2 \rangle \ll 1$.

## III. RESULTS

The shape parameters for the simulation of the nonresonant stellarator divertor are $\varepsilon_0 = 1/2$, $\varepsilon_t = 1/2$, and $\varepsilon_x = -0.31$. The stellarator has five periods, $n_p = 5$. All five periods are considered to be exactly identical. The rotational transform on the magnetic axis of the period is $\iota_0 = 0.15$. The step-size of the map is $\delta\zeta = 2\pi/3600$. The phase portraits of the nonresonant stellarator divertor in the poloidal planes $\zeta = 0$ and $\zeta = \pi$ are shown in Fig. 1. The outermost confining surface is at $r/b = 0.87$, $\theta = 0$, and $\zeta = 0$, where $r/b = \sqrt{\psi_t / \bar{\psi}_g}$. The average normalized toroidal flux inside the outermost confining surface is $\psi_0 = \bar{\psi}_{t,LGS} / \bar{\psi}_g = 1.3429$. $\psi_{t,LGS}$ denotes the toroidal flux inside the outermost confining surface. The wall is circular with radius $r_{wall}/b = 4$. The largest radial excursion of the outermost confining surface is $r/b = 2$. See Fig. 1(c). The rotational transform $\iota$ for the single period is shown in Fig. 2. The good magnetic surface located midway between the magnetic axis and the outermost confining surface is chosen as the starting surface for field lines. The surface is generated by 10 000 toroidal circuits of the period. Every 10th point in the $\zeta = 0$ poloidal plane is chosen as a starting position on this surface. See Fig. 1. These 1000 lines are advanced for 100 000 toroidal circuits of the period in forward and backward directions using the maps for a fixed value of the radial velocity $u_\psi$. The strike points of field lines on the wall are calculated. This procedure is repeated for radial velocity $u_\psi / \bar{\psi}_g = 1, 0.9, 0.8, \ldots,$ $5 \times 10^{-6}$, and $4 \times 10^{-6}$. For each value of the radial velocity, the footprint on the wall is calculated for forward and backward moving lines.

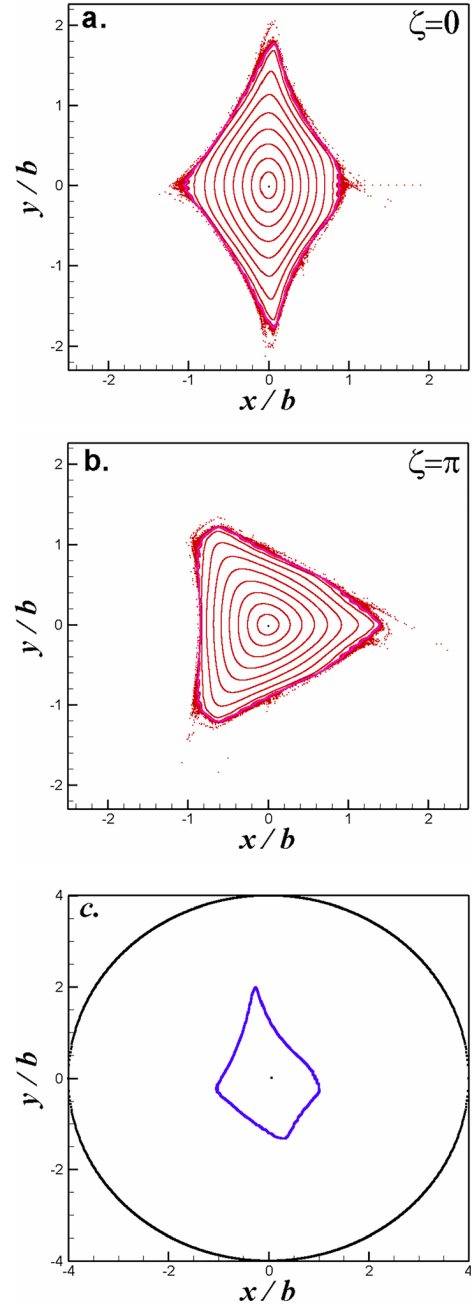

FIG. 1. (a) The phase portrait in the $\zeta = 0$ poloidal plane of a single period in the nonresonant stellarator divertor, (b) the phase portrait in the $\zeta = \pi$ poloidal planes of a single period in the nonresonant stellarator divertor, and (c) the intercepting wall at $r/b = 4$ and the outermost confining surface with the largest radial excursion of $r_{max}/b = 1.9998$, which occurs in the poloidal plane $\zeta = 0.9949$ radians. The intercepting wall is a conformal circular torus with radius $r/b = 4$. Color codes: blue: the outermost confining surface; green: the starting surface midway between the magnetic axis and the outermost confining surface in $r$-space; black = the intercepting wall at $r/b = 4$; blue = the outermost confining surface with the largest radial excursion in the $\zeta = 0.9949$ radians poloidal plane.

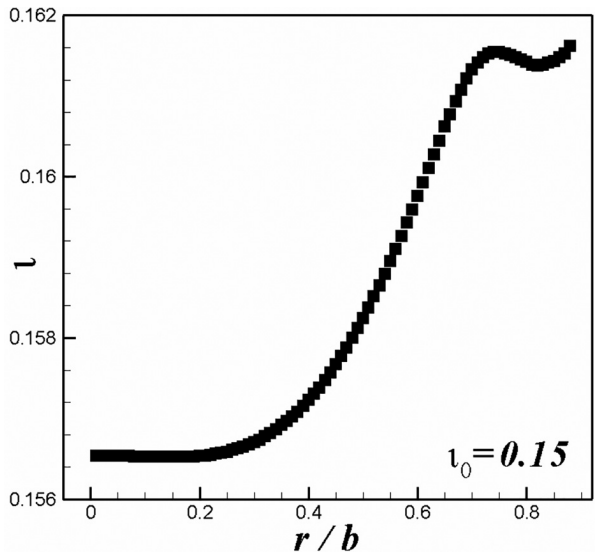

**FIG. 2.** The rotational transform $\iota = d\theta/d\zeta$ for a single period with $\iota_0 = 0.15$.

### A. The topology of the footprints

The most significant features of the footprints found in the study are: (1) the footprints are stellarator symmetric.[9] (2) The location of the footprints on the wall is fixed for all velocities. See Fig. 3. (3) When the radial velocity is large, $u_\psi > 10^{-2}$, the field lines go into both families of magnetic cantori. Field lines go into the primary family for all values of radial velocity, see Figs. 4(a)–4(d). The field lines go into the secondary family when velocities are sufficiently large, $u_\psi > 10^{-2}$, see Figs. 4(a) and 4(b). (4) For all values of velocities, some field lines form a pair of continuous toroidal stripes on the wall. These images are made up of two stripes above and below the turnstiles covering the whole range of $\zeta$ from 0 to $2\pi$. In Figs. 3 and 4, these stripes on the wall are not shown. The nature of these stripes is not adequately understood and will undergo further study. (5) The footprints of the primary family are stellarator symmetric. The footprint of the primary family, which lies mostly in the upper-half $(\zeta, \theta)$ plane, is called the upper primary footprint, which is made up of strike points of the forward lines, see Figs. 3 and 4. The footprint of the primary family, which lies mostly in the lower-half $(\zeta, \theta)$ plane, is called the lower primary footprint. The lower primary footprint is made up of the strike points of the backward lines, see Figs. 3 and 4. The upper and the lower primary footprints are stellarator symmetric. The primary footprints have the largest toroidal extent $\Delta\zeta \sim 9$ radians when $u_\psi = 1$, see Fig. 4(a); and $\Delta\zeta \sim 4.7$ radians when $u_\psi = 10^{-5}$, see Fig. 4(d). (6) Similarly, the footprints of secondary family are also stellarator symmetric. The footprint of the secondary family, which lies in the upper-half $(\zeta, \theta)$ plane, is called the upper secondary footprint. It is made up of the strike points of the backward lines, see Figs. 3 and 4. The footprint of the secondary family, which lies in the lower-half $(\zeta, \theta)$ plane, is called the lower secondary footprint. It is made up of the strike points of the forward lines, see Figs. 3 and 4. The upper and the lower secondary footprints are stellarator symmetric. The secondary footprints have the toroidal extent $\Delta\zeta \sim 4.6$ radians when $u_\psi = 1$, see Fig. 4(a); and $\Delta\zeta \sim 4.5$ radians when $u_\psi = 2 \times 10^{-2}$, see Fig. 4(b). So, the variation in $\Delta\zeta$ for secondary footprints is marginal.

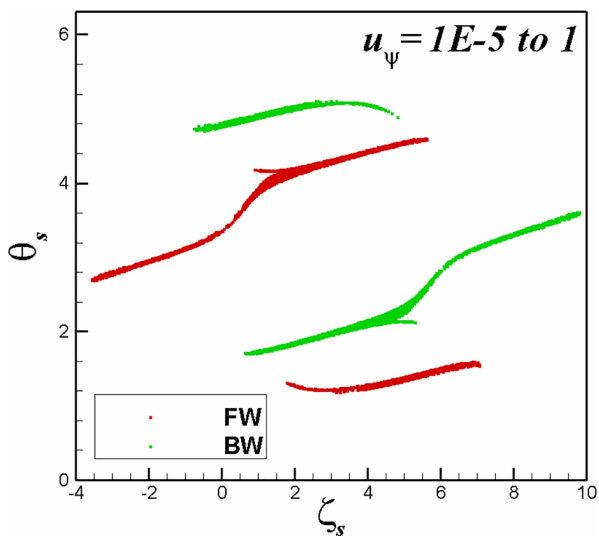

**FIG. 3.** The footprints for radial velocities $u_\psi = 1.0, 0.9, 0.8, \ldots, 3 \times 10^{-5}, 2 \times 10^{-5}, 10^{-5}$. All the footprints are shown together. The footprints are stellarator symmetric. For $u_\psi > 10^{-2}$, the footprints are made up of flux tubes of the first family, flux tubes of the second family, and a pair of the projections of the starting surface on the wall (not shown). For $4 \times 10^{-6} \leq u_\psi \leq 10^{-2}$, the footprints are made of the flux tubes of the primary family and the pair of projections of the starting surface on the wall (not shown). Color codes: red: the strike points of the forward lines, green: the strike points of the backward lines.

### B. The structure of magnetic turnstiles

The structure of the primary and secondary footprints when they coexist for high velocities $u_\psi > 10^{-2}$, and when only the primary footprints exist for small velocities $u_\psi \leq 10^{-2}$ is shown in Figs. 5–7. The footprints of each family are stellarator symmetric. The footprints have sharp boundaries and are hollow. This means the lines enter and travel along the edges of the turnstiles. For the primary footprints when $u_\psi \leq 10^{-2}$, one arm of the footprint is shorter than the other, Figs. 7(a) and 7(b). On the one end, the two arms are joined, and on the other end, they are not joined. Some of these features make it difficult to define and estimate the width of footprints.

### C. Calculation of probability exponents

When the velocities are large, $10^{-2} < u_\psi \leq 1$, the field lines go into the magnetic turnstiles of both the primary and secondary families of cantori; and when the velocities are small, $u_\psi < 10^{-2}$, the field lines go only into the turnstiles of the primary family of cantori. Each family of cantori has two stellarator symmetric turnstiles. The two turnstiles of a family of cantori form a double flux tube. The forward lines go into one tube, and the backward lines go into the other tube. We use the simulation data to estimate the probability exponents $d_1$ and $d_2$ for the primary and the secondary families of magnetic cantori in the nonresonant stellarator divertor.

The widths of the footprints are estimated using the method given in Sec. II E. The width of a footprint is the scatter $\delta\theta$, Eq. (11).

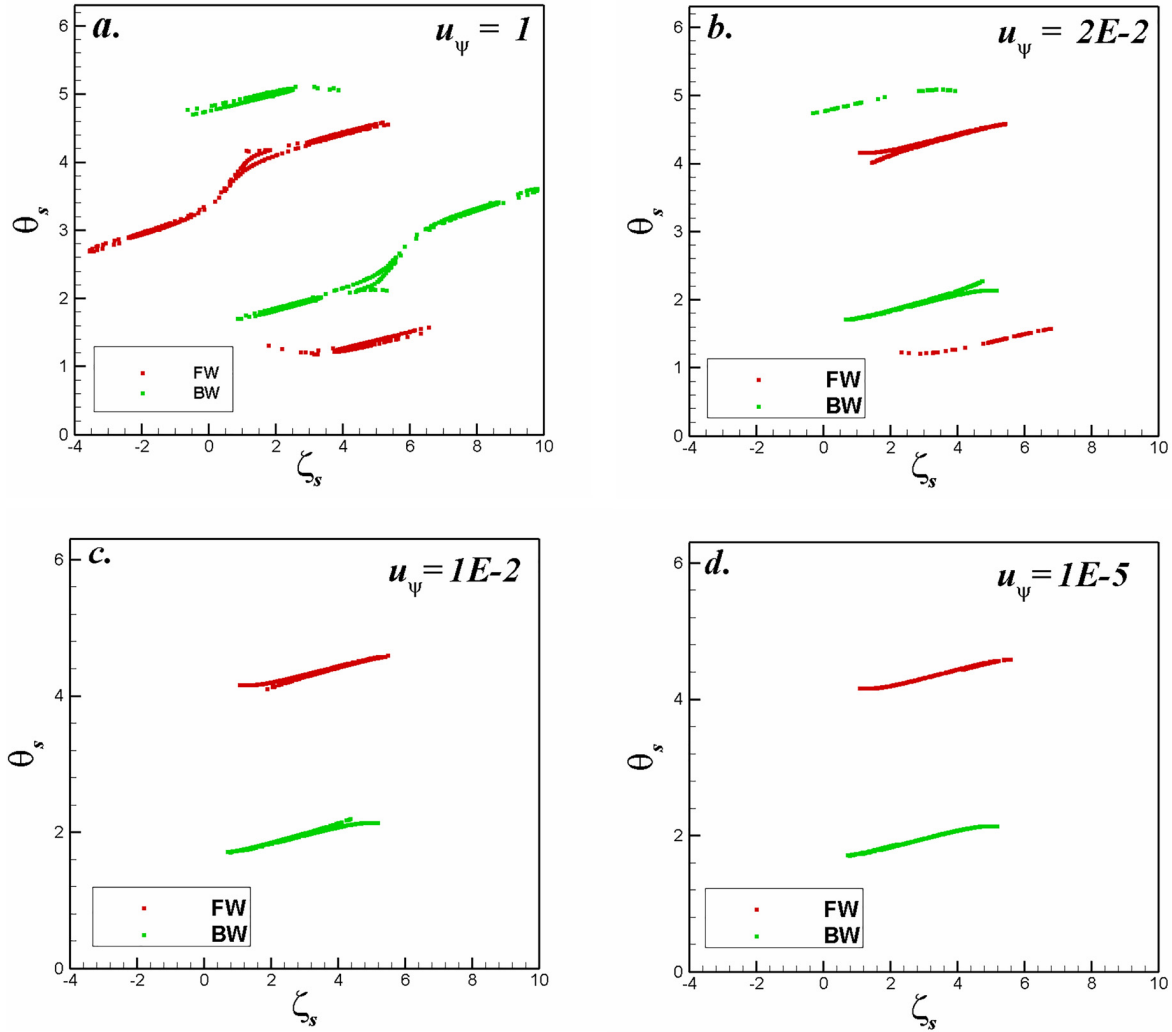

**FIG. 4.** Footprints for (a) $u_\psi = 1$, (b) $u_\psi = 2 \times 10^{-2}$, (c) $u_\psi = 10^{-2}$, and (d) $u_\psi = 10^{-5}$. Color codes: red = footprints of the forward lines, green = footprints of the backward lines. The projections of the starting surface are not shown.

Most of the widths are $w = \delta\theta \sim 0.1$–$0.2$ radians. For a minor radius of 0.5 m, these widths correspond to 5–10 cm. The loss-time $\zeta_l$ is defined as the number of toroidal transits of a single period of the stellarator when the number of field lines remaining in plasma has fallen to $1/e$ of the starting value. The loss-time is calculated from when the first line hits the wall, $\zeta_0$.

### 1. The probability exponent for the primary family of cantori

We use the simulation data for $u_\psi \leq 10^{-2}$ to estimate the probability exponent $d_1$ for the primary family of cantori. In this case, the forward lines go into the upper turnstile, and the backward lines go into the lower turnstile in the $(\zeta, \theta)$ plane; see Fig. 4. The widths of the footprints for the forward and the backward lines as functions of velocity $u_\psi$ are fluctuating and irregular. This could be because of the shape and the structure of the primary footprints, see Fig. 7. The data on widths of the footprints, $w(u_\psi)$, for the primary cantori are unresolvable to make an estimate of the scaling of width with velocity $u_\psi$.

Both the underlying magnetic equilibrium and the wall are stellarator symmetric. So, ideally the widths and the loss-times of the forward and the backward lines for a given velocity must be exactly equal. Departures from the exact equality occur due to statistical and numerical errors in simulation. When the departures are sufficiently small, the simulation data can give us reliable, accurate, and consistent estimates of scalings and the probability exponents. The average width of the footprint for a given velocity is $\langle w \rangle = (w_{FW} + w_{BW})/2$, and the difference between the widths of the forward and backward lines for a given velocity is $\Delta w = |w_{FW} - w_{BW}|$. Here, $w_{FW}$ is the width of footprint of forward lines and $w_{BW}$ is the width of the footprint of the backward lines for a given velocity $u_\psi$. Then, the deviation from exact equality of widths of the forward and the backward lines for a given

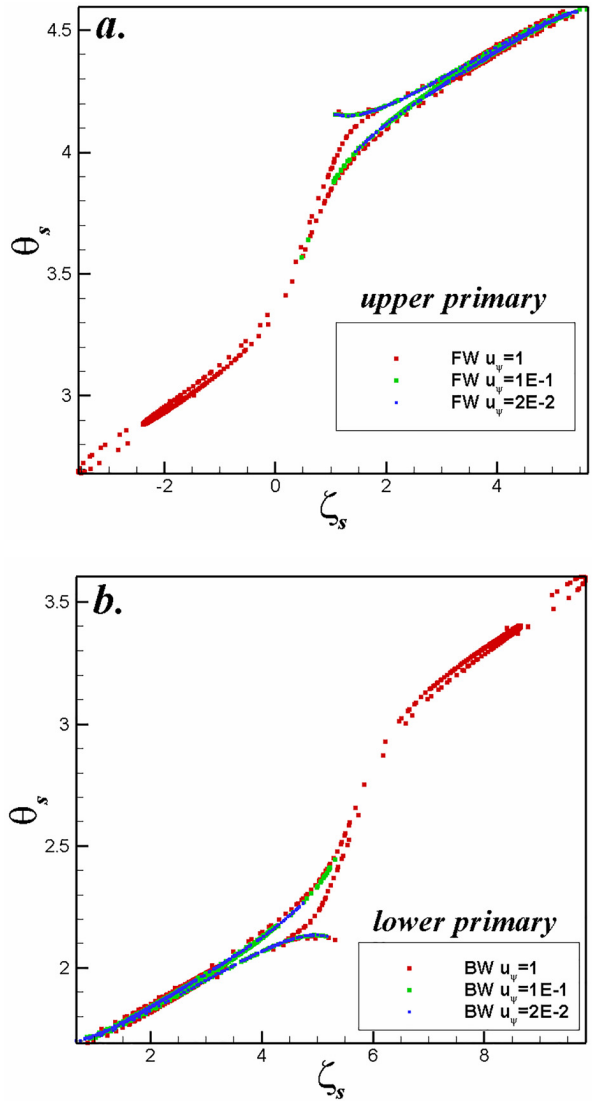

FIG. 5. The primary footprints when $u_\psi \geq 2 \times 10^{-2}$. (a) The upper primary footprints, and (b) the lower primary footprints. Color codes: red = when $u_\psi = 1$; green = when $u_\psi = 0.1$; and blue = when $u_\psi = 2 \times 10^{-2}$.

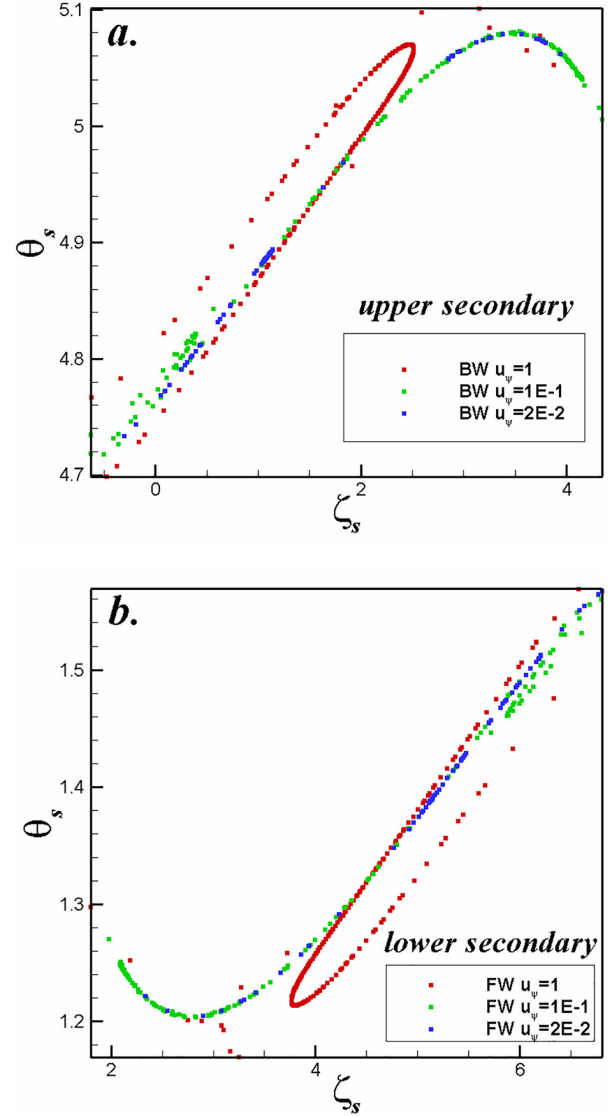

FIG. 6. The secondary footprints when $u_\psi \geq 2 \times 10^{-2}$. (a) The upper secondary footprints, and (b) the lower secondary footprints. Color codes: red = when $u_\psi = 1$; green = when $u_\psi = 0.1$; and blue = when $u_\psi = 2 \times 10^{-2}$.

velocity is $\delta w = \Delta w / \langle w \rangle$. Similarly, the deviation of the loss-times from exact equality of the loss-times of the forward and the backward lines for a given velocity is $\delta \zeta_l = \Delta \zeta_l / \langle \zeta_l \rangle$. When the lines go only into the primary cantori, there are only two values of the velocity for which $\delta w < 2\%$ and there are eight values of velocity for which the deviation $\delta \zeta_l < 2\%$. These data points with $\delta \zeta_l < 2\%$ are used to estimate the scaling of the loss-time with velocity for the primary cantori, see Fig. 8. Linear fit to $\log[\langle \zeta_l \rangle (u_\psi)]$ vs $\log(u_\psi)$ gives

$$\zeta_l = c u_\psi^p, \quad (12)$$

where $c = 0.8857$, $0.6898 < c < 1.1372$, $p = -0.6921 \pm 0.0280$, and the coefficient of multiple determination for the fit is $R^2 = 0.9903$. From Eq. (12), the probability exponent $d_1$, Eq. (8), for the primary family is

$d_1 = 2.2480$ and the error in $d_1$ is $1.9774 < d_1 < 2.5728$. From this, $d_1 = 2.480 \cong 2\frac{1}{4} = \frac{9}{4}$. Our previous value[6] of $d_1 = 2$ is within the error bar of $d_1$ calculated here.

### 2. The probability exponent for the secondary family of cantori

We use the simulation data for large velocities, $10^{-2} < u_\psi \leq 1$, to estimate the probability exponent $d_2$ for the secondary family of cantori. For these large velocities, we increase the number of field lines from 1000 to 100 000. We make the increment in the velocities 10 times smaller, $\Delta u_\psi = 0.01$. So, $u_\psi = 1.00, 0.99, 0.98, \ldots, 0.02$. For large velocities, the simulation data in the range $0.37 < u_\psi \leq 1$, are regular;

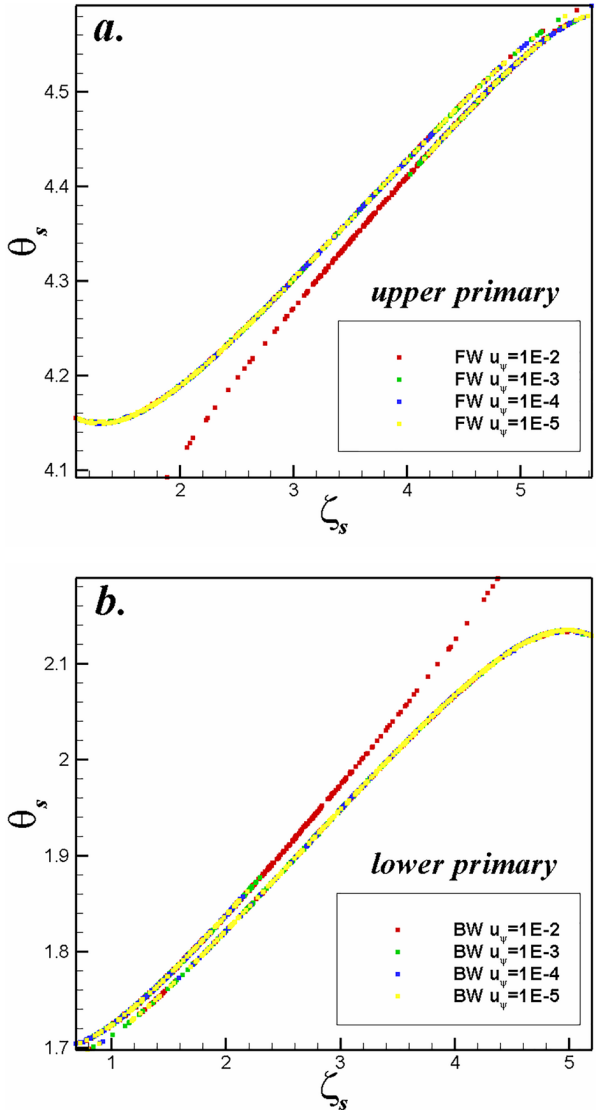

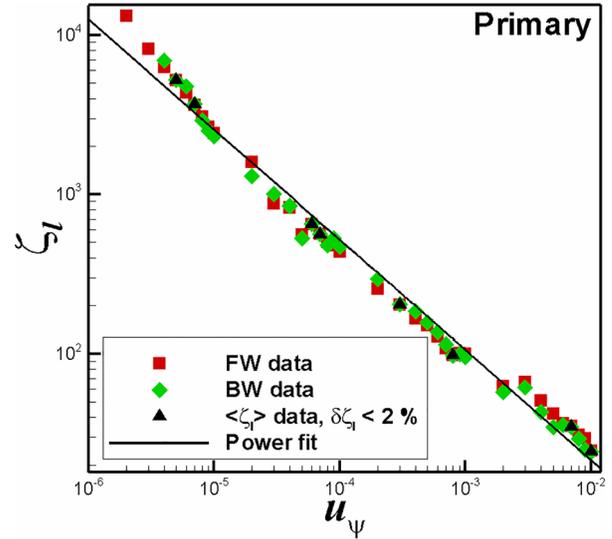

**FIG. 8.** The scaling of the loss-time $\zeta_l$ with the velocity $u_\psi$ for the primary family of cantori. Color code: red squares = the loss time for the forward lines, green diamonds = the loss-time for the backward lines, black triangles = the average loss-time $\langle \zeta_l \rangle$ when the deviation of loss-times from exact equality $\delta\zeta_l < 2\%$. The loss-time $\zeta_l$ scales as $1/u_\psi^{0.6921}$, giving the probability exponent $d_1 = 2.2480$.

**FIG. 7.** The primary footprints when $u_\psi \leq 10^{-2}$. (a) The upper primary footprints, and (b) the lower primary footprints. Color codes: red = when $u_\psi = 10^{-2}$; green = when $u_\psi = 10^{-3}$; blue = when $u_\psi = 10^{-4}$, and yellow = when $u_\psi = 10^{-5}$.

the widths of the secondary footprints are monotonously decreasing functions of the velocity and the loss-times are monotonously increasing functions of velocity, and the departures of the widths and the loss-times from exact equalities are small. The forward lines go into the lower turnstile, and the backward lines go into the upper turnstile in the $(\zeta, \theta)$ plane; see Fig. 4. We use the simulation data in the range $0.37 < u_\psi \leq 1$ to estimate $d_2$. The data on widths and loss-times in the range $10^{-2} < u_\psi < 0.37$ are fluctuating and irregular. This could be because the system is making a phase transition from going into the cantori of both families to going into the cantori of only primary family.

For $0.37 < u_\psi \leq 1$, there are 50 data points for which the deviation $\delta w < 1\%$ and all the 64 data points have $\delta\zeta_l < 1\%$, see Figs. 9 and 10. For $u_\psi$ close to 1, the deviation $\delta w > 1\%$; $\delta w$ becomes larger as $u_\psi \to 1^-$, see Fig. 9. We use these 50 datapoints for the average widths and the 64 data points for the average loss-times to calculate the scalings of the width and loss-time with velocity. A consistent, reliable, and accurate estimate of $d_2$ is obtained for the scaling laws of the form $w \propto 1/(u_\psi - C)^2$ and $\zeta_l \propto 1/u_\psi^3$.

The width $w$ scales as

$$w = A + \frac{B}{(u_\psi - C)^2}, \quad (13)$$

where $A = 0.0380 \pm 0.0003$, $B = 0.0038 \pm 0.0001$, $C = 1.0643 \pm 0.0029$, and $0.37 \leq u_\psi \leq 1$. $(u_\psi - C)$ is a shifted velocity. The coefficient of multiple determination for the fit is $R^2 = 0.9987$. Figure 9(a) shows the simulation data and the scaling given by Eq. (13). A linear fit to log-log plot of the data gives the slope of the line $m = -1.9937 \pm 0.0076$; see Fig. 9(b). Then, from Eq. (7), the probability exponent is $d_2 = -1.5016$. From the linear fit to the log-log plot, the error in $d_2$ is $-1.5035 < d_2 < -1.4997$. The error in $d_2$ is 0.2531%.

The loss-time scales as

$$\zeta_l = D + \frac{E}{u_\psi^3}, \quad (14)$$

where $D = 0.7797 \pm 0.0036$ and $E = 0.0399 \pm 0.0005$. The coefficient of multiple determination is $R^2 = 0.9896$. Figure 10(a) shows the data and the scaling given by Eq. (14). Linear fit to log-log plot of the simulation data gives the slope $m = -3.1424 \pm 0.0582$, see Fig. 10(b). Then, from Eq. (8), the probability exponent is $d_2 = -1.4668$. The error in $d_2$ is $-1.4798 < d_2 < -1.4544$. The error in $d_2$ is 1.7317%.

We have two estimates for $d_2$, $-1.5035$ and $-1.4668$. The correlation coefficient $R^2$ for the fit that gave the value $-1.5035$ is larger than that for $-1.4668$; and the error for $-1.5035$ is much smaller than that for $-1.4668$. So, the value $-1.5035$ is more reliable and more accurate. Therefore, we choose $d_2 = -1.5035 \cong -3/2$.

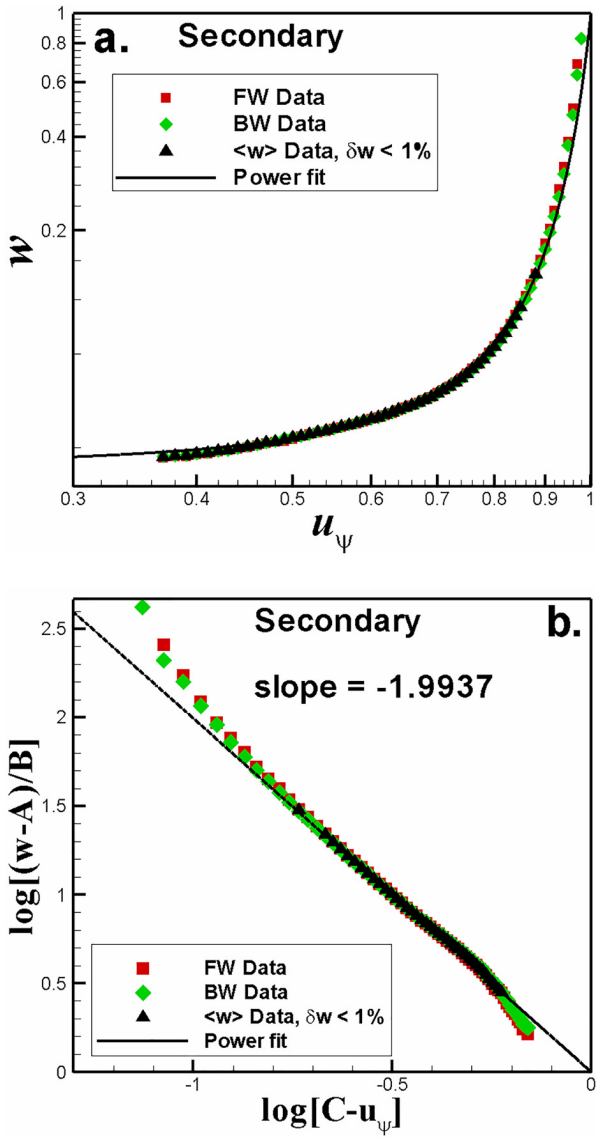

FIG. 9. (a) The power law fit for the width of the footprints of the secondary family of cantori, Eq. (13). (b) Log-log plot of (a). Color code: red squares = widths of the footprints of forward lines, green diamonds = widths of the footprints of the backward lines, black triangles = the average widths $\langle w \rangle$ when the deviation $\delta w < 1\%$, Black curve = the power law fit, Eq. (13), to $\langle w \rangle$ as a function of velocity $u_\psi$.

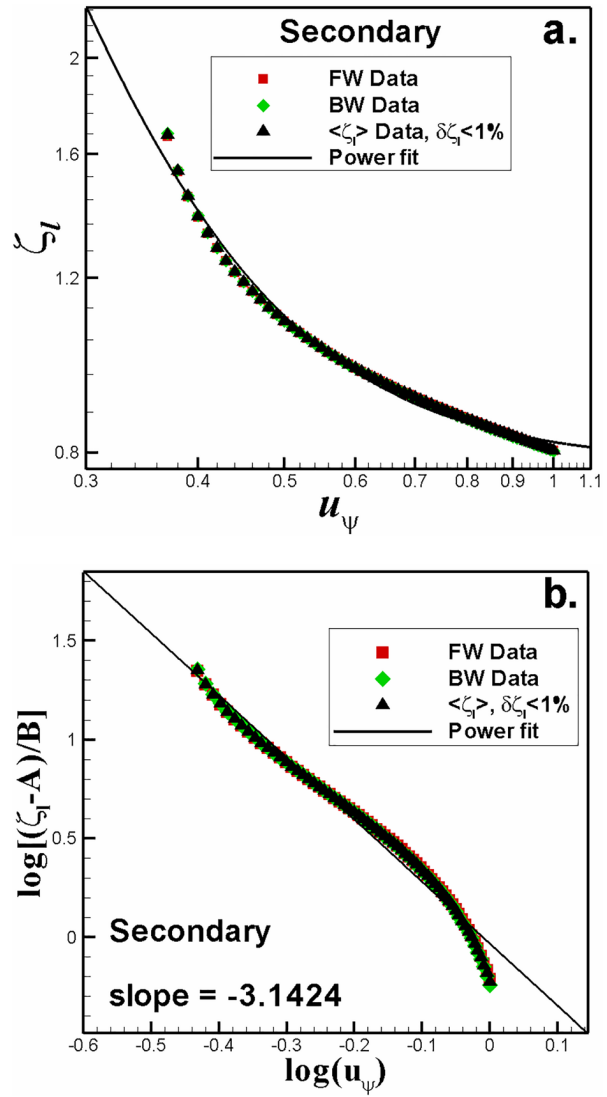

FIG. 10. (a) The power law fit, Eq. (14), for the loss-times of the secondary family of cantori. (b) Log-log plot of the (a). Color code: red squares = the loss-times for the forward lines, green diamonds = the loss-times for the backward lines, black triangle = the average loss-times $\langle \zeta_l \rangle$ when the deviation $\delta \zeta_l < 1\%$, black curve = the power law fit, Eq. (14), to $\langle \zeta_l \rangle$ as a function of velocity $u_\psi$.

The key result from the simulation is the calculation of the probability exponents

$$d_1 = 2\frac{1}{4} \quad (15)$$

and

$$d_2 = -1\frac{1}{2}. \quad (16)$$

Because $d_1 \neq d_2$, the scaling is not universal. It is interesting to note that since $d_1 = 9/4$ and $d_2 = -3/2$, $d_2 = -\sqrt{d_1}$.

Generally, probability exponents are expected to be positive. So, the negative probability exponent $d_2$ is unexpected. This can possibly be understood in the context of a terminal radial velocity $u_\psi$ such that for velocities smaller than the terminal velocity, no lines go into the secondary turnstiles. However, the scaling of the width and the loss-time resulting from the negative exponent are physical and expected. The nonuniversality of the exponents, $d_1 \neq d_2$, is not unexpected. There is no compelling physics reason for the scalings to be universal.

The deviation of the log-log slopes from perfectly fitting a straight line (the black line) is only in part statistics. There is no reason why the power law approximation in the theory must be exact. Indeed, there are good reasons to expect it not to be—the whole of turnstile theory depends on multiple high order islands and their properties. Therefore, it is remarkable that the power law approximation works so well.

## IV. SUMMARY, CONCLUSIONS, AND DISCUSSION

The nonresonant stellarator divertor with strictly identical periods is studied using the simulation method developed in Ref. 6. The outermost confining surface has sharp edges, and the magnetic surfaces are elliptical in the poloidal plane $\zeta = 0$ and triangular in the poloidal plane $\zeta = \pi$. The field lines are given a velocity in $u_\psi$ in $\psi_t$-space to explore the magnetic topology. The velocity $u_\psi$ is varied over roughly six orders of magnitude from $u_\psi/\psi_0 = 1$ to $4 \times 10^{-6}$. The field lines start on a good surface midway between the outermost confining surface and the magnetic axis. The field lines with radial velocity are integrated in both the forward and the backward directions using a symplectic map. They first explore the magnetic topology of the good magnetic surfaces, exit the outermost surface, and explore the magnetic topology between the outermost surface and the wall, and then strike the wall.

The region between the outermost confining surface and the wall has two families of cantori—the primary and the secondary family. The magnetic turnstile of each family consists of a double flux tube. Outgoing lines exit through one tube, and incoming lines enter through the other. The intersections of the tubes with the wall are the footprints. Lines go into both families when the velocities are high, $10^{-2} < u_\psi \leq 1$; and the lines go only into the primary family when the velocities are low, $4 \times 10^{-6} \leq u_\psi \leq 10^{-2}$. Because the underlying magnetic equilibrium and the wall are both stellarator symmetric, the footprints of each family are also stellarator symmetric for all values of velocity. The location of the footprints on the wall is fixed for all velocities. As the velocity becomes smaller, the footprint shrinks in size.

Some lines form a pair of continuous toroidal stripes on the wall. Forward lines go into one stripe and backward lines go in the other stripe. These stripes are also stellarator symmetric.

The widths and the loss-times of the footprints are calculated as functions of the velocity. For the primary family, the data on widths of footprints are unresolvable to estimate the scaling of width with velocity. Since both the magnetic equilibrium and the wall are stellarator symmetric, the widths and loss-times of the footprints of the forward lines and the backward lines must be exactly equal for a given velocity. This fact is used to get the best estimates of the probability exponents. The loss-time of primary family scales as $-0.6921$ power of velocity. From this, the best estimate of the exponent $d_1$ is 9/4 with the error $1.9774 < d_1 < 2.5728$. For the secondary family, the best estimate of the exponent $d_2$ is $-3/2$ with the error $-1.5035 < d_2 < -1.4997$. The scalings are not universal.

Robust location of the footprints on the wall is a highly desirable feature of the divertor. It is consistent with the assertion that the divertor flux tubes tend to have a robust strike location on walls independent of plasma details.[2] It is also consistent with the study by Bader et al. on the HSX.[2] It is remarkable that good estimates of exponents are possible despite the very complicated nature of magnetic cantori with high order islands.

The key findings of this study are as follows: (1) there are two families of magnetic cantori—the primary and the secondary. For large spiraling velocities, field lines go into both families, and for small spiraling velocities, lines go only into the primary family. (2) When the underlying magnetic equilibrium and the wall are both stellarator symmetric, the footprints of a family are also stellarator symmetric. (3) The probability exponent of the primary family is $d_1 = 9/4$; and the secondary family is $d_2 = -3/2$. (4) The location of the footprints on the wall is fixed.

It will also be important to study if the robustness of the locations of footprints on the wall holds when the underlying equilibrium is changed, or when the stellarator symmetry is violated, or when the periodicity breaks.

## ACKNOWLEDGMENTS

This material was based upon work supported by the U.S. Department of Energy, Office of Science, Office of Fusion Energy Sciences under Award Nos. DEFG02-03ER54696 to Columbia University and DE-SC0020107 to Hampton University. This research used resources of the NERSC, supported by the Office of Science, U.S. DOE, under Contract No. DE-AC02-05CH11231. The authors thank the reviewer for valuable comments and suggestions.